\documentclass[aps,pre,amsmath,amsfonts,amssymb,floatfix,superscriptaddress,
nofootinbib]{revtex4}
\usepackage{mathrsfs} \usepackage{graphicx,color} \usepackage{bbm}
\usepackage{multirow}
\usepackage{amsmath}
\usepackage{amssymb}
\usepackage{algorithm,algpseudocode}

\usepackage{rotate}
\usepackage{epsfig}
\usepackage{psfrag}
\usepackage{mathtools}
\usepackage{float}

 \let\b=\beta \let\g=\gamma \let\d=\delta
  \let\h=\eta

   \let\G=\Gamma

\def\PP{{\cal P}}\def\EE{{\cal E}}

\def\to{\rightarrow}

\newcommand{\beq}{\begin{equation}} \newcommand{\eeq}{\end{equation}}

\newcommand{\qq}[1]{[\![{#1}]\!]}

\newcommand\be{\begin{equation}}
\newcommand\bea{\begin{eqnarray} \nonumber }
\newcommand\ee{\end{equation}}
\newcommand\eea{\end{eqnarray}}

\begin{document}

\title{Optimal Inflation Target: \\
Insights from an Agent-Based Model}

\author{Jean-Philippe Bouchaud}
\affiliation{
CFM, 23 rue de l'Universit\'e, 75007 Paris, France
}

\author{Stanislao Gualdi}
\affiliation{
CFM, 23 rue de l'Universit\'e, 75007 Paris, France
}

\author{Marco Tarzia}
\affiliation{
Universit\'e Pierre et Marie Curie - Paris 6, Laboratoire de Physique
Th\'eorique de la Mati\`ere 
Condens\'ee, 4, Place Jussieu, 
Tour 12, 75252 Paris Cedex 05, France }

\author{Francesco Zamponi}
\affiliation{Laboratoire de physique th\'eorique, D\'epartement de physique de
l'ENS, \'Ecole normale sup\'erieure, PSL
Research University, Sorbonne Universit\'es, UPMC Univ. Paris 06, CNRS,
75005 Paris, France}

\begin{abstract} 
Which level of inflation should Central Banks be targeting? We investigate this issue in the context of a simplified Agent Based Model of the economy. 
Depending on the value of the parameters that describe the behaviour of agents (in particular inflation anticipations), we find a rich variety 
of behaviour at the macro-level. {Without any active monetary policy, our ABM economy can be in a high inflation/high output state, or in a low inflation/low output state. 
Hyper-inflation, deflation and ``business cycles'' between coexisting states are also found.}
We then introduce a Central Bank with a Taylor rule-based inflation target, and study the resulting aggregate variables. 
Our main result is that too-low inflation targets are in general detrimental to a CB-monitored economy. One symptom is a persistent under-realisation of
inflation, perhaps similar to the current macroeconomic situation. Higher inflation targets are found to improve both unemployment and negative interest rate episodes. 
Our results are compared with the predictions of the standard DSGE model. 
\end{abstract} 

\maketitle

\tableofcontents

\clearpage

\section{Introduction}

Most Central Banks around the world nowadays adjust their monetary policy to
reach a 2\%/year inflation target. The rationale for choosing 2\% rather than
1\% or 3\% is however not clear, 
as with many other ``magic numbers'' religiously used in economic policy. The
recent crisis has put to the fore the problem of negative nominal interest
rates, which can be seen as a consequence of
low inflation targets and thus low baseline rates. As
emphasized by O.~Blanchard in 2010~\cite{Blanch10}, {\it ``As a matter of logic, higher average
inflation and thus higher average nominal 
interest rates before the crisis would have given more room for monetary policy
to be eased during the crisis.''}
This view is however disputed by many
economists, who strongly argue 
against a raise of the inflation target (see e.g.~\cite{Ball,Bernanke} for a
recent overview, {and for a discussion of the historical origin of the 2\% target}). A major argument to that effect is the credibility of Central
Banks, who have succeeded in anchoring low-inflation 
expectations in the minds of economic agents. If the inflation target is changed
in the face of new circumstances, these expectations may un-moor, and 
the very efficiency of monetary policy may suffer as a consequence. Clearly, the
fear of a lurking run-away inflation is weighing heavily on the debate. 

Yet, the question of an ``optimal'' inflation level is well worth considering,
and policy makers are eager to receive inputs from academic research. As Federal
Reserve Chairwoman J. Yellen recently declared~\cite{Yellen1}: {\it 
``We very much look forward to seeing research by economists that will help
inform our future decisions on this''.} Of course, optimality needs to be
defined and different criteria (i.e. welfare functions) 
may lead to different results. More important still is the modelling framework
used to describe the economy. A clear puzzle is that standard monetary theories
imply
zero or negative optimal inflation rates, at variance with Central Banks'
inflation targets~\cite{Uribe}. The standard DSGE machinery -- the ``workhorse'' of
monetary economists
~\cite{DSGE} -- 
has recently been extended to cope with non-zero inflation rates, and generally
concludes that the optimal inflation rate should be smaller than
2\%~\cite{Coibion}. However, DSGE models
are based on a series of highly debatable assumptions, and have been under
intense fire after the 2007 crisis: see the insightful set of contributions in 
the Oxford Review of Economic Policy~\cite{OREP}; see also~\cite{Kirman,Buiter,Solow,Bookstaber}.

Another route is provided by Agent Based Models (ABM) in which reasonable
behavioural rules replace the representative DSGE agent with a fully rational
long-term plan. ABMs can include a number of economically relevant features which would be very
difficult to accommodate within the DSGE
straight-jacket~\cite{Eurace,SantAnna,Lagom,Review,Haldane}. 
Many simplifying, sometimes ad-hoc assumptions are of course necessary, but a
considerable advantage of ABMs is that interaction-induced, collective effects
are present, 
whereas DSGE models reduce the whole economy to a small number of representative
agents. As a consequence, the global ``equilibrium'' state of the economy 
is an emergent property in the former case, while it is a {\it deus ex machina}
in the latter case. 

In particular, {\it crises} (i.e. large swings in the output) can occur
endogenously within ABMs~\cite{DelliGatti,Tipping}. DSGE models, on the other
hand, only describe small, mean-reverting fluctuations around the postulated
equilibrium and
crises can only result from exogenous, unpredictable shocks.\footnote{R. Lucas
famously argued that the 2008 crisis was not predicted because economic theory
predicts that such events cannot be predicted~\cite{Lucas}. {However, as discussed in~\cite{OREP}, a benchmark macroeconomic
model should be able to account for crises because it is crises that have the largest effects on individual well-being.}} As a case in point,
we found in~\cite{Policy} that the aggregate {behavior} of the economy is not a
smooth function of the baseline interest rate: the fact that firms are risk
averse and fear going into debt leads to more unemployment that can spiral into
a {destabilizing} feedback loop. This is one of the ``dark corners''~\cite{Blanchard} that ABMs can help uncovering.

{While there is a growing literature on monetary policies in ABMs~\cite{SantAnna,Salle,Salle2,Anufriev,Arifovic,Grauwe},
the optimal inflation target question has not been investigated using ABMs (although see~\cite{Salle} where the success of {\it targeting policies} is discussed 
within the framework of an ABM).} In this paper, we take on
this issue using an arguably over-simplified, bare-bone ABM dubbed ``Mark-0'', 
studied in great details in~\cite{Tipping,Policy} following previous work
by the group of Delli Gatti et al.~\cite{DelliGatti} (see
also~\cite{Lengnick2,Salle}). As discussed in~\cite{Tipping}, the Mark-0 economy can
be in different ``macro-states'' (HIHO: high inflation/high input, or LILO: low
inflation/low output), depending on various parameters of the model. 
These parameters describe in a phenomenological
way the behaviour of agents (firms, households and banks), and their response to
different economic stimuli. Interestingly, small changes in the value of these parameters
can indeed induce sharp variations in aggregate output, unemployment or
inflation~\cite{Tipping,Policy}.\footnote{{In fact, as will be shown below, HIHO and LILO states can even 
coexist for some parameter range, see Figure \ref{fig:native} below. This coexistence region was overlooked in an earlier version of this paper, 
and accounts for some of the puzzling effects that we initially reported.}} This allows us to consider 
different baseline economies  and study the influence of the chosen inflation
target on the total output, on the real interest rate and 
on the probability of negative nominal interest rates.

Our main conclusion is that in general, increasing the inflation target reduces
unemployment and reduces the probability of negative rates. 
Unsurprisingly, it also reduces real interest rates on savings. In fact, trying
to impose low inflation on an economy that would naturally run at full steam
with high inflation can lead to an output collapse. On the other hand, high
inflation policies can be dangerous and may generate hyper-inflation if agents 
lose faith in the ability of Central Banks to fulfill their mandate.

\section{Methodological remarks and scope of the paper} \label{sec:method}

Before going into the technical discussion about the model and the results, 
we want to stress here some methodological aspects, and state a few disclaimers.

\begin{itemize}

\item
As recalled above, we are fully aware that our stylized ABM relies on somewhat 
arbitrary assumptions and is unrealistic on several counts. We have discussed in detail 
the logic of our approach in~\cite{Tipping}. While micro-rules should {\it in fine} be 
justified by direct empirical data on the behaviour of households and firms, many results 
are in fact generic and robust against changes of these micro-rules -- allowing one to 
draw important qualitative conclusions from such stylized models. It thus seems to us more relevant, 
at this stage of the ABM research agenda, to stick with the now well-studied Mark-0 model 
and explore the issue of inflation target in a ``proof of concept'' manner. A number of possible 
improvements are listed in the conclusion of the present paper.

\item
Although far from perfect, Mark-0 contains plausible ingredients that are most probably present in reality. 
For example, our model encodes in a schematic manner the consumption {behavior} of households facing inflation, that is in fact
similar to the standard Euler equation for consumption in general equilibrium
models~\cite{DSGE}. We also account for the effect of inflation on the policy of indebted firms, which
appears to be absent in DSGE models.  The fact that our results strongly contrast with those of standard DSGE models is in our opinion 
enough to motivate in-depth investigations of more realistic ABMs, and more empirical work
on the micro/behavioural assumptions that underpin these models. 

\item
Our approach is not normative, in the sense that we do not consider any 
specific welfare function that should be maximized. Our aim is rather to provide a synthetic ``dashboard'' of the 
simulated economies, with inflation, unemployment, probability of negative rates and real rates on deposits, as
a function of the target inflation level. Although not formalized, it will be clear from these dashboards that some 
inflation targets are qualitatively better than others. Remaining at the level of qualitative statements seems to us a 
way to avoid the ``pretense of knowledge'' syndrom~\cite{Caballero,Buiter}. As Keynes said, {\it it is better to be roughly right than exactly wrong.}

\item
The Mark-0 model can exhibit very different behaviours depending on the parameters~\cite{Tipping}, with regions where the economy collapses. It is important to stress 
that the behavioral rules that define our model are supposed to be reasonable when the economy behaves normally, and we believe that they correctly describe how such a normal 
state can become unstable. However, once the instability happens and the economy truly collapses, it is of course unreasonable 
to expect that agents will keep acting according to the same rules, in particular the Central Bank and other institutions. However, this is besides the point we want to make in this paper.

\item
It is somehow unavoidable that complex ABMs can lead to non-intuitive results, in particular concerning the detailed shape of the phase diagrams, 
for example the appearance of a region of parameters where good and bad states of the economy can coexist. Although it might seem unsatisfactory not to
have a fully analytical understanding of these transitions, we claim that it is in fact one of the strength of ABMs: to be able to elicit scenarios that are hard to imagine 
for the unaided human mind, because they are the result of non-linearities and sometimes antagonist feedback loops. 

\item
Finally, an important disclaimer: the parameters chosen in the following are not the result of a precise
calibration. We only made reasonable guesses in order to have reasonable numbers as outputs of the model (e.g. reasonable values of yearly inflation). 
All the numbers quoted below are not intended to be taken literally (although we believe they should be taken seriously!).
\end{itemize}

\section{A short recap on Mark-0}
\label{sec:mark0recap}

The Mark-0 model with a Central Bank (CB) and interest rates has been described
in full details in~\cite{Tipping,Policy}, where pseudo-codes are also provided.
We will not repeat here the full logic of the model, but only focus on the
elements that are relevant for determining inflation in the three sectors:
households, firms and the CB. The pseudo-code of Mark-0, {where all the modelling choices are made explicit}, 
can be found in Appendix~\ref{app:Mark0}.  

First, we need some basic notions. The model is defined in discrete time, where the
unit time between $t$ and $t+1$ is plausibly of the order of months. For
definiteness, 
we will choose in the following the unit time scale to be 6 months. Each firm
$i$ at time $t$ produces a quantity $Y_i(t)$ of perishable goods
that it attempts to sell at price $p_i(t)$, and pays a wage $W_i(t)$ to its
employees. The demand $D_i(t)$ for good $i$ depends on the global consumption
budget of households $C_B(t)$, 
itself determined as an inflation rate-dependent fraction of the household
savings. $D_i$ is a decreasing function of the firm price $p_i$, with a price
sensitivity 
parameter that can be tuned. To update their production, price and wage policy,
firms use reasonable ``rules of thumb''~\cite{Tipping} that also depend on the
inflation rate through 
their level of debt (see below). For example, production is decreased and
employees are made redundant whenever $Y_i > D_i$, and vice-versa.\footnote{As a
consequence 
of these adaptive adjustments, the economy is on average always `close' to the
global market clearing condition one would posit in a fully representative agent
framework. 
However, small fluctuations persists in the limit of large system sizes giving
rise to a rich phenomenology~\cite{Tipping}, including business cycles.} The
model is fully 
``stock-flow consistent'' (i.e. all the stocks and flows within the toy economy
are properly accounted for).

The instantaneous inflation rate $\pi(t)$ is defined as: 
\be
\pi(t) = \frac{\overline p(t) - \overline p(t-1)}{\overline p(t-1)}; \qquad
\overline p(t) = \frac{\sum_i p_i(t) Y_i(t)}{\sum_i Y_i(t)},
\ee
where $\overline p(t)$ is the production-weighted average price. We will assume
that firms, households and the CB do not react to the instantaneous value of
$\pi(t)$, but rather 
to a smoothed, exponential moving average $\pi^{\text{ema}}(t)$ of the realised inflation, computed as 
\be
\pi^{\text{ema}}(t) = \omega \pi(t) + (1-\omega)\pi^{\text{ema}}(t-1),
\ee
where we fix $\omega=0.2$, which corresponds to an averaging time of $\approx
4.5$ time steps, i.e. roughly 2 years in our setting. {Note that all quantities noted with the superscript ``ema'' in the following 
are defined in the same way, with the same numerical value of $\omega$.}

In Mark-0 we assume a linear production function with a constant unit productivity, which means that output and
employment coincide. The unemployment rate $u$ is defined as:
\be
u(t) = 1 - \frac{\sum_i  Y_i(t)}{N},
\ee
where $N$ is the number of firms, which also coincides with the total workforce~\cite{Tipping}. {Note that firms 
cannot hire more workers than available, so that $u(t) \geq 0$ at all times -- see Eq. (\ref{y_update}) below.} 

\subsection{The Central Bank policy}

In this work, for simplicity we consider a single-mandate CB that attempts to steer the economy
towards a target inflation level $\pi^\star$ (in~\cite{Policy}, we in fact
considered a double-mandate CB
also targeting a certain employment level $\varepsilon^\star$). The monetary
policy followed by the CB for fixing the base interest rate is described by a 
Taylor-like rule of the form~\cite{Mankiw,DSGE}:\footnote{{The original Taylor rule reads \cite{Taylor}: $\rho_0(t) = \rho^\star + \pi^{\text{ema}}(t) + \phi'_\pi \, [\pi^{\text{ema}}(t)-{\pi}^\star]$ 
which amounts to the substitution $\rho^\star \to \rho^\star + \pi^\star$ and $\phi_\pi \to 1 + \phi'_\pi$ in Eq. (\ref{taylor_rule}).}} 
\be
\label{taylor_rule}
\rho_0(t) = \rho^\star + \phi_\pi \, [\pi^{\text{ema}}(t)-{\pi}^\star] 
\ee
where $\rho^\star$ is the {baseline interest rate} and $\phi_{\pi}>0$ quantifies
the intensity of the policy.\footnote{Note that this is the only action taken by the CB to achieve the target; in particular, no
actions on the quantity of circulating money, such as quantitative easing or printing money can be taken by the CB.} {In the following, we do not impose a Zero Lower Bound to
the base-line interest rate, in view of the recent monetary policy; imposing it does not affect substantially our conclusions.
Note that here $\rho^\star$ cannot be interpreted as the ``natural'' interest rate, which is itself an emergent property of the model, that depends on all
the parameters. In this sense, $\rho^\star$ is another parameter of the model, that contributes to the determination of the macroeconomic state
(see Fig.~\ref{fig:native_ema} below).
}
 
We assume that the banking sector -- described at the aggregate level by 
a single ``representative bank'' -- 
sets the interest rates on deposits and loans ($\rho^d(t)$ and $\rho^\ell(t)$ respectively) uniformly for all lenders and borrowers\footnote{
{This is, in our model, the only role played by the banking sector: a transmission belt of the CB policy. In reality, the banking sector has much more freedom, and can sometimes make the 
CB policy ineffective, e.g. by restricting credit even in presence of a strong incentive from the CB.}}. Therefore, the rate $\rho^\ell$
increases and $\rho^d$ decreases when the firm default rate increases, in such a way that the banking sector {(i.e., the representative bank)} -- which fully absorbs these defaults -- 
makes zero profit {at each time step} (see~\cite{Policy} {and the pseudo-code provided in Appendix \ref{app:Mark0}} for more details). 

\subsection{Households}

The effect of inflation on households is the standard trade-off between {saving} (at rate $\rho^d$) and consumption.  
We therefore assume that the total consumption budget of households $C_B(t)$ is
given by:
\be
\label{cons_budget}
C_B(t) = c(t) \left[ S(t) + W(t) + \rho^d(t)S(t) \right]\quad \text{with} \quad
c(t)=\qq{ c_0 \left[1+\alpha_c (\widehat{\pi}(t)-\rho^{d,\text{ema}}(t))\right] } \ ,
\ee
where $S(t)$ is the savings, $W(t)$ the total wages, $\widehat{\pi}(t)$ is the
{\it expected} inflation in the next period -- see Eq. (\ref{exp-infl}) below --
and 
$c(t)$ is the consumption propensity, which is clipped to the interval $[0,1]$.
This is expressed by the symbol $\qq{x}$ which
means that the quantity $x$ is boxed between $0$ and $1$, i.e. $\qq{x}=1$ if $x>1$, $\qq{x}=0$ if $x<0$,
and $\qq{x}=x$ otherwise.
This propensity 
is equal to {a baseline value} $c_0$ when the difference between expected inflation and the
interest paid on their savings is zero, and increases (decreases) when this
difference is positive (negative). The parameter $\alpha_c>0$ 
determines the sensitivity of households to the real interest rate. Eq. \eqref{cons_budget} {describes a feedback of inflation on consumption} similar to the standard Euler equation of DSGE models
(see e.g.~\cite{Mankiw,DSGE}). {The total household savings evolve according to:
\be
S(t+1) = S(t) + W(t) + \rho^d(t)S(t) - C(t),
\ee
where $C(t) \leq C_B(t)$ is the actual consumption of households, see~\cite{Tipping}.}

We furthermore posit that the expected inflation $\widehat{\pi}(t)$ is given by
a linear combination of the realised inflation $\pi^{\text{ema}}(t)$ and the CB
target inflation $\pi^\star$ (see also \cite{Salle}):
\be\label{exp-infl}
\widehat{\pi}(t) = \tau^{\text{R}} \, \pi^{\text{ema}}(t) + \tau^{\text{T}} \,
\pi^\star.
\ee
The parameters $\tau^{\text{R}}$ and $\tau^{\text{T}}$ (``R'' for realised and ``T'' for target) can be interpreted as
capturing {the importance of past inflation and} the trust of economic agents in the ability of the CB to enforce its
inflation target. When $\tau^{\text{R}}=0$ and $\tau^{\text{T}}=1$,
agents fully trust that the target inflation will be realised. When
$\tau^{\text{R}} > 0$, they are also influenced by the past realised inflation
when they form their expectations. When $\tau^{\text{R}} > 1$, they expect 
more inflation to be realised in the next period. As we will see below, this can give rise to
hyper-inflation episodes, which is the scenario that prevents (in
the mind of many policy makers and of the public opinion) higher inflation targets.

{In principle, $\tau^{\text{R}}$ and $\tau^{\text{T}}$ should depend on the commitment of the 
Central Bank, captured by the parameter $\phi_\pi$. 
In particular, the function $\tau^{\text{T}}(\phi_\pi)$ should be,
for consistency, such that $\tau^{\text{T}}=0$ when $\phi_\pi=0$. 
In fact, in the absence of an active CB ($\phi_\pi=0$), one should assume that the
inflation expectation parameter $\tau^{\text{T}}$ is zero, since there is no anchoring force to a definite inflation target.
Although we do not introduce a precise model for
$\tau^{\text{T}}(\phi_\pi)$, below we always assume that $\tau^{\text{T}}=0$ when $\phi_\pi=0$.} Also, $\tau^{\text{R}}$ and $\tau^{\text{T}}$ might 
be time dependent, as economic agents compare the realised inflation to the target inflation and ``learn'' about the
credibility of the CB -- see below and \cite{Salle} for a discussion of this particular point. In the present 
paper, we will treat $\tau^{\text{R}}$ and $\tau^{\text{T}}$ as time independent, {leaving this interesting development for future work.}

\subsection{Firms}

\subsubsection{Financial fragility} 

The model contains $N_{\rm F}$ firms,
each firm being characterized by its production $Y_i$ (equal to its workforce in our zero growth economy),
demand for its goods $D_i$, price $p_i$, wage $W_i$ and its cash balance $\EE_i$
which, when negative, is
the debt of the firm. We characterize the {\it financial fragility} of the firm
through the debt-to-payroll ratio
\be
\Phi_i=-\frac{\EE_i}{W_i Y_i}.
\ee
{Negative $\Phi$'s describe healthy firms with positive cash balance, while indebted firms have a positive $\Phi$.} 
If $\Phi_i < \Theta$, i.e. when the flux of credit needed from the bank is
not too high compared to the size of the company (measured as the total payroll), the firm $i$ is allowed
to continue
its activity. If on the other hand $\Phi_i \geq \Theta$, the firm $i$
defaults and the corresponding default cost is absorbed by the banking sector, 
{which adjusts the loan and deposit rates $\rho^\ell$ and $\rho^d$ accordingly. The defaulted firm is replaced by a new one, initialised at random (using the 
average parameters of other firms).}
The parameter $\Theta$ controls the maximum leverage in the economy,
and models the risk-control policy of the 
banking sector.

\subsubsection{Production update} 

If the firm is allowed to continue its business, it adapts its price, wages and
production according to
reasonable (but of course debatable) ``rules of thumb'' -- see~\cite{Tipping,Policy}. In particular, the
production update is chosen as:
 \beq
 \label{y_update}
\begin{split}
    \text{If } Y_i(t) < D_i(t)  &\hskip10pt \Rightarrow \hskip10pt 
     Y_i(t+1)=Y_i(t)+ \min\{ \eta^+_i ( D_i(t)-Y_i(t)), u^\star_i(t) \} \\
    \text{If }   Y_i(t) > D_i(t)  &\hskip10pt  \Rightarrow \hskip10pt
    Y_i(t+1)= Y_i(t) - \eta^-_i [Y_i(t)-D_i(t)]  \\
\end{split}
\eeq
where $u^\star_i(t)$ is the maximum number of unemployed workers available to
the firm $i$ at time $t$ (see~\cite[Appendix A]{Policy}). 
The coefficients $\eta^\pm \in [0,1]$ express the sensitivity of the firm's
target production to excess 
demand/supply. We postulate that the production adjustment depends on the
financial fragility $\Phi_i$ of the firm:
firms that are close to bankruptcy are arguably faster to fire and slower to
hire, and vice-versa for healthy firms. In order to
model this tendency, we posit that the coefficients $\eta^\pm_i$ for firm $i$ (belonging to $[0,1]$)
are given by: 
\bea
\eta^-_i = \qq{\eta_0^- \, (1+ \Gamma \Phi_i(t))} \\
\eta^+_i = \qq{\eta_0^+ \, (1- \Gamma \Phi_i(t))}, 
\eea 
where $\eta_0^\pm$ are fixed coefficients, identical for all firms.
The factor $\Gamma > 0$ measures how the financial fragility of firms influences
their hiring/firing policy, 
since a larger value of $\Phi_i$ then leads to a faster downward adjustment of
the workforce 
when the firm is over-producing, and a slower (more cautious) upward adjustment
when the firm is under-producing.

In~\cite{Policy} we argued that $\Gamma$ should in fact depend on the difference
between the interest rate and the inflation: 
high cost of credit makes firms particularly wary of going into debt and their
sensitivity to their financial 
fragility is increased. Therefore, we postulate that interest rates influence
the firm's policy through the financial fragility sensitivity $\Gamma$, as:
\be
\label{alpha_Gamma}
\Gamma = \max{\{\alpha_\Gamma
({\rho}^{\ell,\text{ema}}(t)-\widehat{\pi}(t)),\G_0\}},
\ee
where $\alpha_\Gamma$ (similarly to $\alpha_c$ above) captures the influence of
the real interest rate on 
loans on the hiring/firing policy of the firms. {This feedback of inflation on firms policy is one of 
the important features of our model.}

\subsubsection{Price update}

{Following the initial specification of the Mark series of models \cite{DelliGatti},} prices are updated through a random multiplicative process which takes into
account the production-demand gap experienced in the previous time step and  
  if the price offered is competitive (with respect to the average price). The
update rule for prices reads:
  \beq
  \label{p_update}
  \begin{split}
    \text{If } Y_i(t) < D_i(t)  &\hskip10pt \Rightarrow \hskip10pt 
    \begin{cases}
&  \text{If } p_i(t) < \overline{p}(t) \hskip10pt \Rightarrow \hskip10pt   
p_i(t+1) = p_i(t) (1 + \g \xi_i(t) )( 1 +  \widehat \pi(t)) \\
&  \text{If } p_i(t) \geq \overline{p}(t) \hskip10pt \Rightarrow \hskip10pt   
p_i(t+1) = p_i(t)( 1 + \widehat \pi(t))  \\
\end{cases}
\\
  \text{If }   Y_i(t) > D_i(t)  &\hskip10pt  \Rightarrow \hskip10pt
  \begin{cases}
&  \text{If } p_i(t) > \overline{p}(t) \hskip10pt \Rightarrow \hskip10pt   
p_i(t+1) = p_i(t) (1 - \g \xi_i(t) )( 1 +  \widehat \pi(t))  \\
&  \text{If } p_i(t) \leq \overline{p}(t) \hskip10pt \Rightarrow \hskip10pt   
p_i(t+1) = p_i(t)( 1 + \widehat \pi(t)) \\
    \end{cases}
\end{split}
\eeq
where $\xi_i(t)$ are independent uniform $U[0,1]$ random variables and $\gamma$
is a parameter setting the relative magnitude of the price adjustment, chosen to be $0.1$
throughout this work. The $(1 + \widehat \pi(t))$ factor implies that 
firms also anticipate inflation when they set their prices. This is precisely
the dreaded self-reflexive mechanism that may lead to hyper-inflation when 
expected future inflation is dominated by past realised inflation (the parameter
$\tau^{\text{R}}$), rather than by the CB inflation target (the parameter
$\tau^{\text{T}}$). 

\subsubsection{Wage update}

The wage update rule follows the choices made for price and
production. {Similarly to workforce adjustments, we posit that} at each time step firm $i$ updates the wage paid to its employees
as:
\beq
\begin{split}
\label{eq:wages}
W^T_i(t+1)=W_i(t)[1+\gamma (1 - \Gamma \Phi_i) (1 - u(t)) \xi^\prime_i(t)][ 1
+ g \widehat \pi(t)]
\quad\mbox{if}\quad
\begin{cases}
Y_i(t) &< D_i(t)\\
\PP_i(t) &> 0 
\end{cases}
 \\
W_i(t+1)=W_i(t)[1-\gamma (1 + \Gamma \Phi_i) u(t) \xi^\prime_i(t)][ 1 + g
\widehat \pi(t)]
\quad\mbox{if}\quad
\begin{cases}
Y_i(t) &> D_i(t)\\
\PP_i(t) &< 0 
\end{cases}
\end{split}
\eeq
where $\PP_i(t)$ is the profit of the firm at time $t$ and $\xi^\prime_i(t)$ an
independent $U[0,1]$ random variable.
If $W^T_i(t+1)$ is such that the profit of firm $i$ at time $t$ with this
amount of wages would have been negative, $W_i(t+1)$ is chosen to be exactly at
the equilibrium point where $\PP_i(t)=0$; otherwise $W_i(t+1) = W^T_i(t+1)$. 
Finally, $g$ is a certain parameter modulating the way wages are indexed to
inflation. We will assume in the following 
full indexation ($g=1$), but choosing $g < 1$ can be useful to stabilize the
Mark-0 economy in periods of hyper-inflation.

Note that within the current model the productivity of workers is not related to their wages. The only channel through which wages 
impact production is that the quantity $u^\star_i(t)$ that appears in Eq.~\eqref{y_update}, which represents the share of unemployed workers accessible
to firm $i$, is an increasing function of $W_i$. Hence, firms that want to produce more (hence hire more) do so by increasing $W_i$, as to attract more 
applicants (see~\cite[Appendix A]{Policy} for details).

{
The above rules are meant to capture the fact that deeply indebted firms seek to reduce wages more aggressively, 
whereas flourishing firms tend to increase wages more rapidly: 
\begin{itemize}
\item If a firm makes a profit and it has a large demand for its good, it will increase the pay of its workers. 
The pay rise is expected to
be large if the firm is financially healthy 
and/or if unemployment is low because pressure on salaries is high. 
\item Conversely, if the firm makes a loss and has a low demand for its good, it will
attempt to reduce the wages. This reduction is more drastic if the company is close to bankruptcy, and/or if
unemployment is high, because pressure on salaries is then low. 
\item In all other cases, wages are not updated.
\end{itemize}}

\subsection{Parameters of the model}

\begin{table}[t]
\begin{tabular}{| l | c | c |}
\hline
Number of firms & $N_{\rm F}$ & 10000 \\
Baseline propensity to consume & $c_0$ & 0.5 \\
Intensity of choice parameter & $\beta$ & 2 \\
Baseline price adjustment parameter & $\gamma$ & 0.1 \\
Baseline firing propensity & $\eta^0_-$ & 0.2 \\
Baseline hiring propensity & $\eta^0_+ $ & $R\eta^0_-$ \\
Fraction of dividends & $\delta$ & 0.02 \\
Bankruptcy threshold & $\Theta$ & 3 \\
Frequency of firm revival & $\varphi$  & 0.1 \\
Share of bankruptcies supported by the firms & $f$ & 0.5 \\
Reaction of consumption to inflation  & $\alpha_c$ & 4 \\
Reaction of firms to interest rates & $\alpha_\Gamma$ & 50 \\
Baseline financial fragility sensitivity & $\Gamma_0$ &  0 \\ 
 Exponentially moving average (ema) parameter & $\omega$ & 0.2 \\
 Indexation of wages to inflation & $g$ & 1  \\
 \hline
\end{tabular}
\caption{Parameters of the Mark 0 model that are kept fixed throughout this work, together with their symbol and value.}
\label{tab:param}
\end{table}

The model, as presented above, has several free parameters, that are specified at the beginning of the pseudo-code
presented in Appendix A.
Some values are fixed throughout this work, using values that have been found in previous work to yield reasonable results~\cite{Tipping, Policy}:
their list is given in Table~\ref{tab:param}.

The parameters that remain to be specified, and whose value is varied in this work to explore the corresponding phase diagram, are
the ratio of hiring/firing propensities $R$, the Taylor rule parameters
$\phi_\pi$, $\pi^*$ and $\rho^{\star}$, and the inflation expectation parameters $\tau^{\text{R}}$ and $\tau^{\text{T}}$.

\section{The ``native'' state of the economy}
\label{results}

In~\cite{Tipping,Policy}, we have shown that the Mark-0 economy, once set in
motion, can settle in a variety of stationary macro-states, where the aggregate
variables behave very differently. The strength of Agent Based
modelling is precisely to show that very
different macro-states can emerge from very similar micro-rules, as parameters are varied.
We will not repeat such an analysis in full here, but focus on the role of a few
variables, relevant to the topic of this paper. We start by analyzing
the case where the CB does not react to inflation (i.e. the Taylor rule Eq.
\eqref{taylor_rule} is with $\phi_\pi=0$ and, as a consequence, $\tau^{\text{T}}=0$). We will see later how a Taylor-rule
based policy of the CB allows it to steer the economy towards a target level of inflation, and when such a policy fails.

\begin{figure}[b]
\includegraphics[scale=0.5]{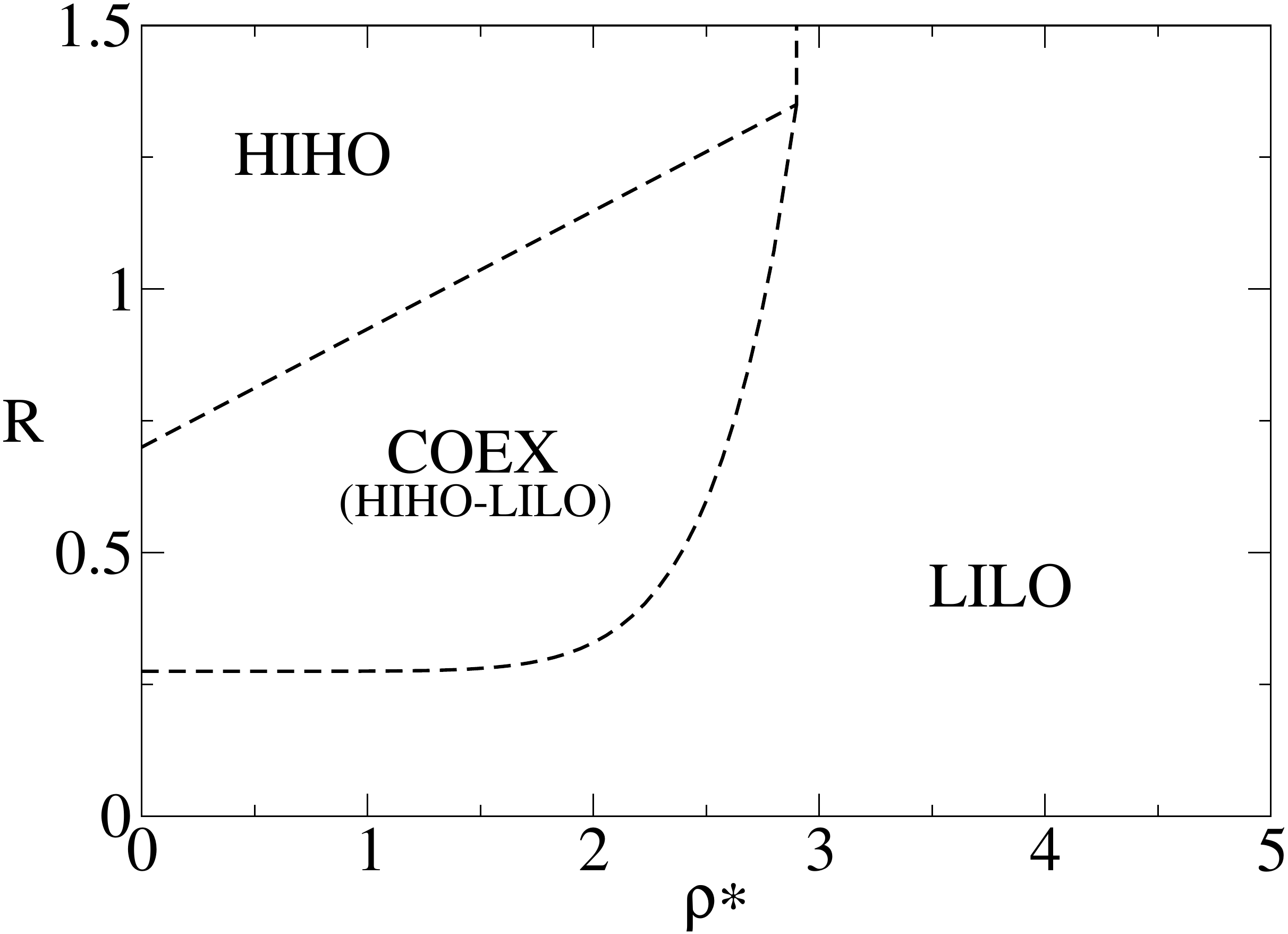}
\caption{
Phase diagram of the model in the $(\rho^{\star},R)$
plane, with $\tau^{\text{R}}=0.5$ and inactive central bank ($\tau^{\text{T}}=0$ and $\phi_\pi=0$). The HIHO phase in the top region of the graph 
is separated from the LILO phase in the bottom region by a discontinuous transition line $\rho^\dagger(R)$. This transition becomes 
a coexistence region for intermediate values of $R$, where the dynamics is non-ergodic: depending on the initial condition, the system ends up either in the LILO state or 
in the HIHO state.}
\label{fig:native}
\end{figure}

Fig.~\ref{fig:native} shows the phase diagram of the model in the plane $(\rho^\star,R)$, where $\rho^\star$ is the baseline interest rate and $R =
\eta_0^+/\eta_0^-$ is the ratio of the hiring propensity to the firing propensity, that was shown  in Ref.~\cite{Tipping} to play a crucial
role for determining the overall state of the economy. {We indeed see that
for $R \lesssim 0.28$ the economy is in a LILO 
state (low inflation and low output/high unemployment). For sufficiently large $R$ and small $\rho^\star$, the economy is in a HIHO state 
(high inflation and high output/low unemployment) and tips over to a LILO state when $\rho^\star > \rho^{\dagger}(R)$. 
This transition is driven by a drop in household consumption and an increased wariness of firms, induced by high yield 
on savings and high cost of loans, see Eqs.~\eqref{cons_budget} and \eqref{alpha_Gamma}}.
As noted in the Introduction, the HIHO/LILO transition occurs discontinuously while the change of interest rate is continuous. The transition line
$\rho^{\dagger}(R)$ is, as expected, an increasing function of $R$ (the economy is more stable where the hiring rate is larger than
the firing rate); it is also a decreasing function of $\alpha_\Gamma$ since firms refrain from taking loans to continue their business when $\alpha_\Gamma$ increases~\cite{Policy}.
{In the proximity of the phase boundaries, the native state of the economy may thus display endogeneous ``business cycles'' corresponding to jumps between these two states.
 
An interesting observation is the presence of a coexistence region where both the LILO state and the HIHO state can be selected by the dynamics, depending on initial 
conditions.\footnote{In fact, very specific initial conditions can also lead to a near collapse of the economy. The possible coexistence between a good state and a bad state of the economy was in fact discussed recently by Carlin and Soskice~\cite{Carlin}.}
For a large system, the dynamics is {\it non-ergodic} and appears to be permanently trapped in one of these two states.}

\begin{figure}[t]
\includegraphics[scale=0.5]{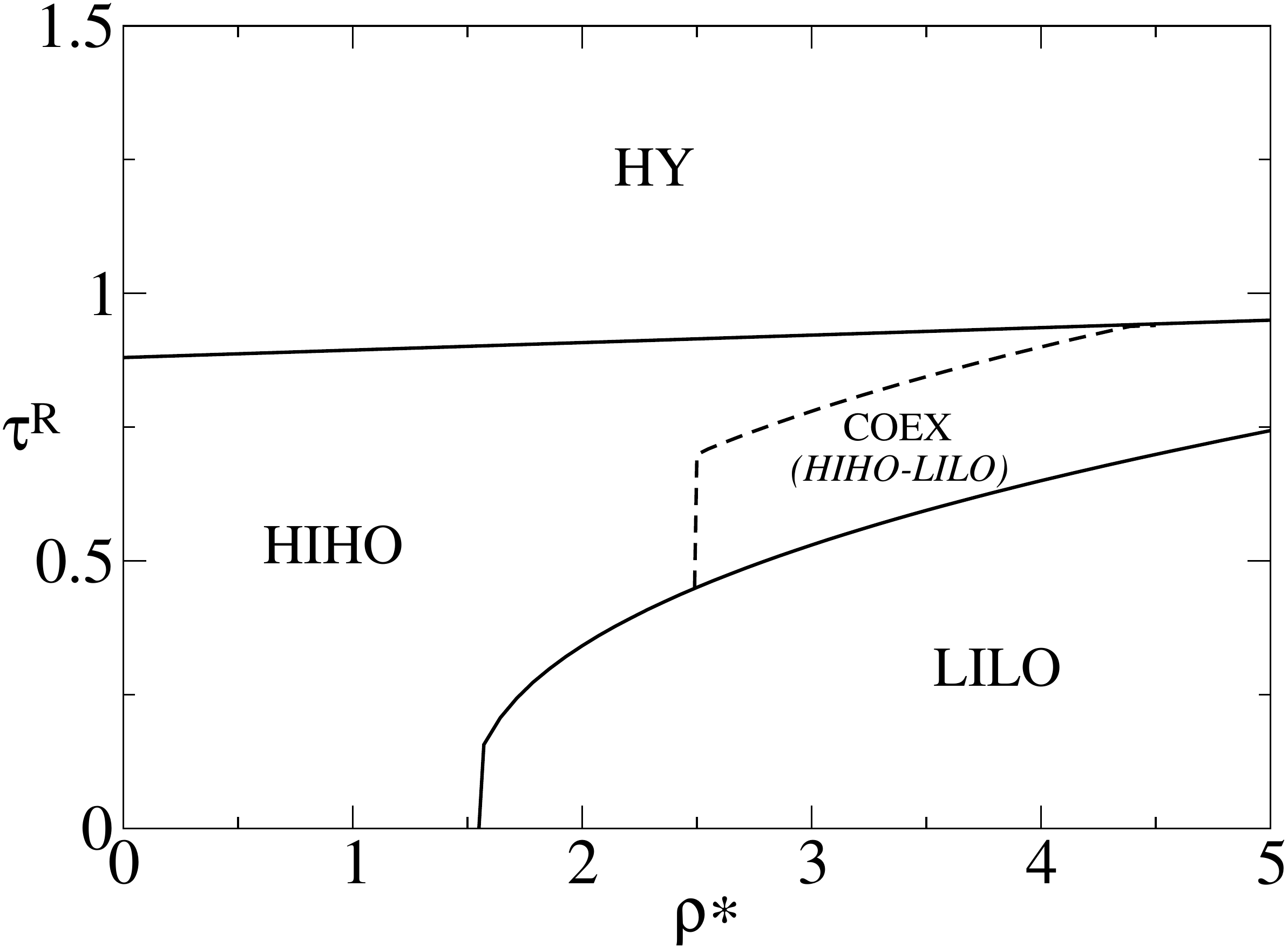}
\caption{
Phase diagram in the $(\rho^{\star},\tau^{\text{R}})$ plane, with $R=1.3$ and inactive central bank ($\tau^{\text{T}}=0$ and $\phi_\pi=0$),
showing a HIHO region, a LILO region and a coexistence region. Note that hyper-inflation is avoided when agents' expectations are sufficiently mean-reverting.
}
\label{fig:native_ema}
\end{figure}

It is interesting to investigate the role of inflation expectations in this
framework, by varying the value of $\tau^{\text{R}}$. Fig.~\ref{fig:native_ema} shows the phase diagram of the model in the plane
$(\rho^\star,\tau^{\text{R}})$ for a fixed value of $R = 1.3$. We find again a HIHO region, a LILO region and a coexistence region for large enough $\rho^\star$ and $\tau^{\text{R}}$. 
{ As anticipated, a transition to a hyper-inflation state (HY) occurs when expectations amplify inflation, more precisely when $\tau^{\text{R}} > \tau^{\dagger} \approx 0.9$. 
The full phase diagram is quite complex, with possible coexistence between hyper-inflation and full employment (HYHO), hyper-inflation and collapse (HYLO) or even hyper-deflation. }

{Note that the HIHO region is characterized, on average, by negative firm profits, and negative real interest rate on deposits
(see Figs.~\ref{fig:targetHIHO} and \ref{fig:targetLILO}).
This might appear counterintuitive, but it happens because our closed economy is characterised by imperfect market clearing (inducing some
waste of goods), and a constant productivity of firms, such that the growth rate of the economy is zero. One could introduce exogenously a
non zero growth rate by allowing productivity to increase as $(1+G)^t$, where $G$ is the growth rate. By slightly amending the rules of the model, this growth rate can be exactly rescaled away by simply shifting all interest rates by $G$, and inflating the money supply by $(1+G)^t$. The real interest rate on deposits, in particular, is shifted by $G$, and the purchasing power of households increases with time. 
} 

\section{Inflation targeting}

We now pick two representative native states of the economy, both for $R=1.3$, in order to be outside of the coexistence region.
The first state 
is with $\rho^\star = 1\%$/year, corresponding to the HIHO state, and the second one is with $\rho^\star = 3\%$/year
corresponding to the LILO state. The inflation level of these native states is, respectively, $4.0 \%$ and $0 \%$
while the unemployment rate is, respectively, $0 \%$ and $39 \%$. 
The long run real return on savings $\langle \rho^{d} - \pi \rangle$ is,
respectively, $-3.0 \%$ and $0 \%$. The HIHO state discourages long term savings
while the LILO state is vastly inefficient in terms of output. 

The CB steps in and modulates the interest rate according to the Taylor rule,
Eq. \eqref{taylor_rule}, with different values of $\phi_\pi$: $1.5,2.5$ and $5$
(the standard value is 
$2.5$, i.e. an increase of inflation by $1\%$ leads to the CB increasing the
nominal base-line rate by $2.5\%$).\footnote{It
would be interesting to extend the present study to dual-mandate CBs.} 
We assume that firms and agents form their inflation expectations by giving an
equal weight to the target inflation $\pi^\star$ and the realised inflation
$\pi^{\text{ema}}$; in other words we set $\tau^{\text{R}}=\tau^{\text{T}}=\frac12$. The results when agents fully trust the CB policy ($\tau^{\text{R}}=0$,
$\tau^{\text{T}}=1$) are not radically different, although, as expected, realised inflation is closer to target. In the other extreme case ($\tau^{\text{R}}=1$, $\tau^{\text{T}}=0$), 
the economy is fraught with instabilities. 

\begin{figure}[b]
\includegraphics[scale=0.5]{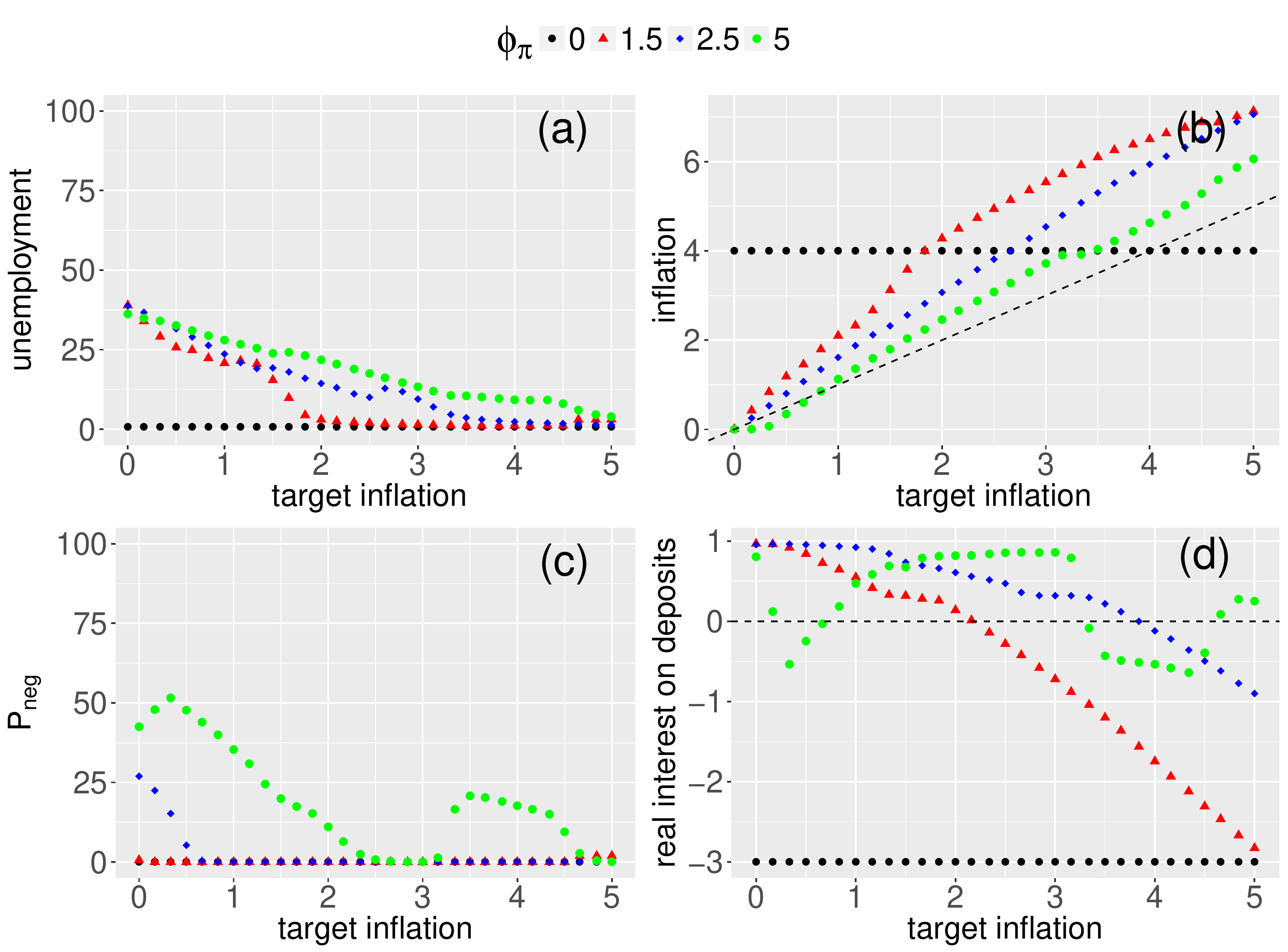}
\caption{
{HIHO native state: Average unemployment (panel a), average realised inflation (panel b), 
probability that the CB must set nominal rates to negative values (panel c)
and average real interest rate paid on deposits (panel d) as a function
of the CB target inflation $\pi^\star$ for 
the native state ($\phi_\pi=0$, $\tau^{\text{T}}=0$, $\tau^{\text{R}}=0.5$, black circles) and for
$\tau^{\text{R}}=\tau^{\text{T}}=0.5$ and $\phi_\pi=1.5$ (red triangles), $\phi_\pi=2.5$ (blue rhombuses),
$\phi_\pi=5$ (green circles). In the last case, large oscillations around these
average values appear, as discussed in~\cite{Policy}. 
Other parameters are: $\rho^{\star}=1\%$ and $R=1.3$. 
Both inflation and rates are expressed as \%/year, unemployment is expressed in \% of the workforce.}}
\label{fig:targetHIHO}
\end{figure}

The resulting states of the monitored economy are summarized in Figs.~\ref{fig:targetHIHO} and~\ref{fig:targetLILO}. In these ``dashboards'' we show, as a function of the inflation target: the average unemployment $\langle u \rangle$, the average realised inflation $\langle \pi \rangle$, the probability $P_{\text{neg}}$ that the CB rate is negative  
and finally the average real interest rate paid on deposits $\langle \rho^{d} -\pi \rangle$, for 
$\phi_\pi=1.5, 2.5$ and $5$.\footnote{{Note however that for $\phi_\pi=5$, there
are large oscillations (``business cycles'') around these average values. As
discussed in~\cite{Policy}, an aggressive CB policy can destabilise the economy.}}

{Starting from the coexistence region (say $R=0.8$) leads to a more complicated discussion, where history plays a role and where economies may be fragile to small perturbations. 
We do not investigate further this intriguing possibility here, but we note that if bistability were to occur in real situations, it would probably urge a radical rethinking of macroeconomic policy.} 

\subsection{High inflation/high output (HIHO) native state}

Starting from a HIHO native state, one sees that targeting a low inflation rate by increasing $\rho_0(t)$ according to the Taylor-rule has a destabilizing effect on output. 
Unemployment rockets to $40 \%$, while realised inflation is indeed low and on target (see Fig.~\ref{fig:targetHIHO}, panel a). 
For our particular ``from-the-hip'' choice of parameters, realised inflation significantly overshoots target when $\pi^\star \gtrsim 1\%$, but this allows unemployment, 
and the probability of negative rates, to be significantly reduced. For example, for $\phi_\pi=2.5$, $P_{\text{neg.}}$ plummets from $\approx 0.25$ when $\pi^\star=0.25 \%$ to zero when
$\pi^\star > 1\%$.

However, the problem with inflation targeting in the HIHO case is that unemployement only falls below $10 \%$ when the 
inflation target is above $3 \%$, in which case the realised inflation is already larger than the natural state inflation of $4 \%$!
In other words, as the CB increases the base interest rate in order to control inflation, it drives the natural HIHO state into a monitored LILO state.
Hence the CB finds it very difficult to achieve simultaneously low inflation and high output. The only upside of the CB policy is that the real interest rate 
on deposits goes from significantly negative ($- 3.0 \%$) in the native HIHO state to positive when $\pi^\star \lesssim 3.5 \%$ in the monitored economy.

The situation improves slightly when agents fully trust the ability of the CB to reach its target (i.e. $\tau^{\text{T}}=1$ and $\tau^{\text{R}}=0$). Hence, stronger anchoring of inflation expectations is beneficial in our ABM setting, in agreement with the intuition gained from DSGE models. At variance with DSGE models, however, large Taylor coefficients (e.g. $\phi_\pi=5$) may lead to instabilities, see \cite{Policy}, and increases significantly the probability of negative nominal rates, see Fig.~\ref{fig:targetHIHO}, panel c).  

\subsection{Low inflation/low output (LILO) native state}

Let us now assume that the underlying economic mechanisms (as described by the
parameters of Mark 0) are such that the native state of the economy is LILO, for example when $R$ is small (firms are more reluctant to hire than to fire) or 
when $\rho^\star$ or $\alpha_\Gamma$ are large (firms are reluctant to take loans). In this case, the role of the CB is to kick start the economy by lowering the interest rate. 

\begin{figure}[!htb]
\includegraphics[scale=0.5]{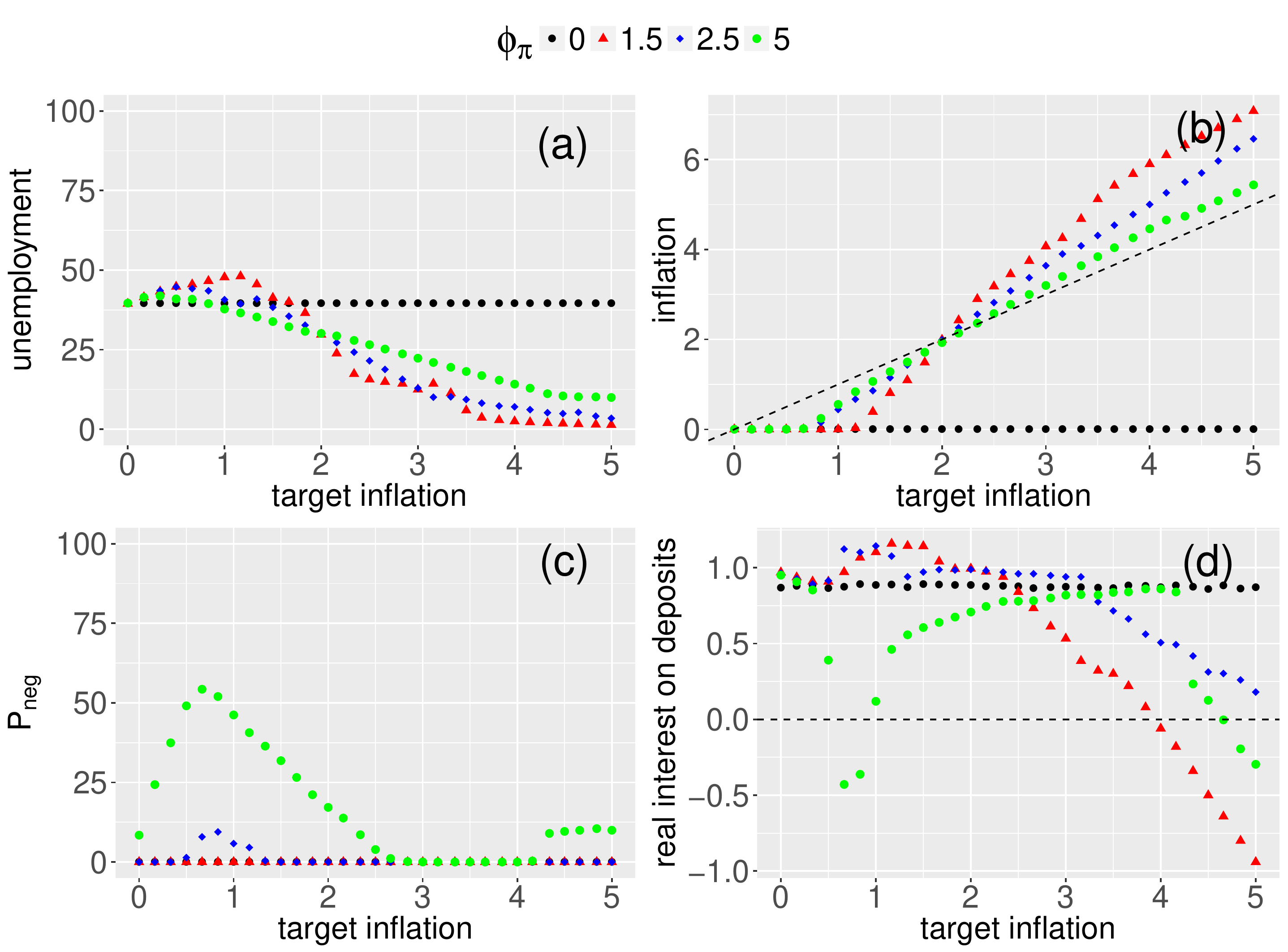}\quad
\caption{{LILO native state: Average unemployment (panel a), average realised inflation (panel b), 
probability that the CB must set nominal rates to negative values (panel c)
and average real interest rate paid on deposits (panel d) as a function
of the CB target inflation $\pi^\star$ for 
the native state ($\phi_\pi=0$, $\tau^{\text{T}}=0$, $\tau^{\text{R}}=0.5$, black circles) and for
$\tau^{\text{R}}=\tau^{\text{T}}=0.5$ and $\phi_\pi=1.5$ (red triangles), $\phi_\pi=2.5$ (blue rhombuses),
$\phi_\pi=5$ (green circles). In the last case, large oscillations around these
average values appear, as discussed in~\cite{Policy}. 
Other parameters are: $\rho^{\star}=3\%$ and $R=1.3$. Both inflation and rates are expressed as \%/year,
unemployment is expressed in \% of the workforce. }}
\label{fig:targetLILO}
\end{figure}

{The results of a Taylor-rule based policy are shown in Fig.~\ref{fig:targetLILO} as a function of the inflation target $\pi^\star$,
again for $\phi_\pi=1.5, 2.5$ and $5$, and $\tau^{\text{T}}=\tau^{\text{R}}=0.5$. 
Surprisingly, the dependence of $\langle u \rangle$ on $\pi^\star$ is found to be {\it non-monotonic}. For $\phi_\pi=2.5$ and $0 < \pi^\star \leq 1.0 \%$, 
unemployment is in fact {\it higher} than in the native state, while realised inflation is {below} target.

Unemployment only dips below $10\%$ when $\pi^\star$ is large enough. For example, when $\pi^\star = 4 \%$, 
unemployment is around $7.5 \%$ (down from $40 \%$ in the native state), long term real savings rate is $0.5 \%$ and the probability of negative nominal rates is zero. 
Realised inflation is however above target: $\langle \pi \rangle \approx 5\%$. }

Fig.~\ref{fig:targetLILO} clearly shows that a symptom of a too-low target is under-realisation of inflation. 
Hence, although quantitative values should not be taken at face value, our analysis suggests that whenever 
$\langle \pi \rangle < \pi^\star$ it is likely that $\pi^\star$ is too low, in the sense that output can be increased by increasing the inflation target.

\subsection{Economic interpretation}

The transmission channels that are responsible for the positive impact of inflation in our model are the following.
First of all, from the update equation for the consumption budget, Eq.~(\ref{cons_budget}), it is clear that stirring the economy 
to a state with higher inflation increases the agents' propensity to consume, thereby boosting demand. 
Perhaps more importantly, increased inflation reduces the real interest rate on loans, and thus the sensivity of firms on their financial fragility: indebted firms are less likely to 
fire and more likely to hire when inflation is larger. Similarly, higher inflation is favourable to the wage policy, see Eq.~(\ref{eq:wages}). 
These effects result in a decrease of the unemployment level and a higher demand that generates a feedback loop that stabilizes the economy and increases the total output.

In a sense, this is the classical ``Keynesian'' positive feedback between the increase of consumption $\to$ increase of output $\to$ decrease of unemployment.
A clear indication that such mechanism is at play is that whenever the CB is successful in stabilizing the economy, reducing the unemployment, 
and increasing the total output, we always find that the realised inflation is larger than $\pi^\star$, due to the fact that the economy tends to amplify 
the effect of the CB. As we will argue below, this feedback is absent in DSGE models. 

Another general argument which allows to rationalize our results is based on our previous observation that the bad states of the economy of the Mark-0 model 
and of its generalizations are often associated with a large amount of ``inactive'' money, stored in the agents' and firms' savings~\cite{Tipping,Policy}.
Increasing the inflation rate induces an erosion of savings and an increased demand, thereby increasing the total amount of money circulating in the
economy.

\subsection{Phillips Curves}

\begin{figure}[!htb]
\includegraphics[scale=0.3]{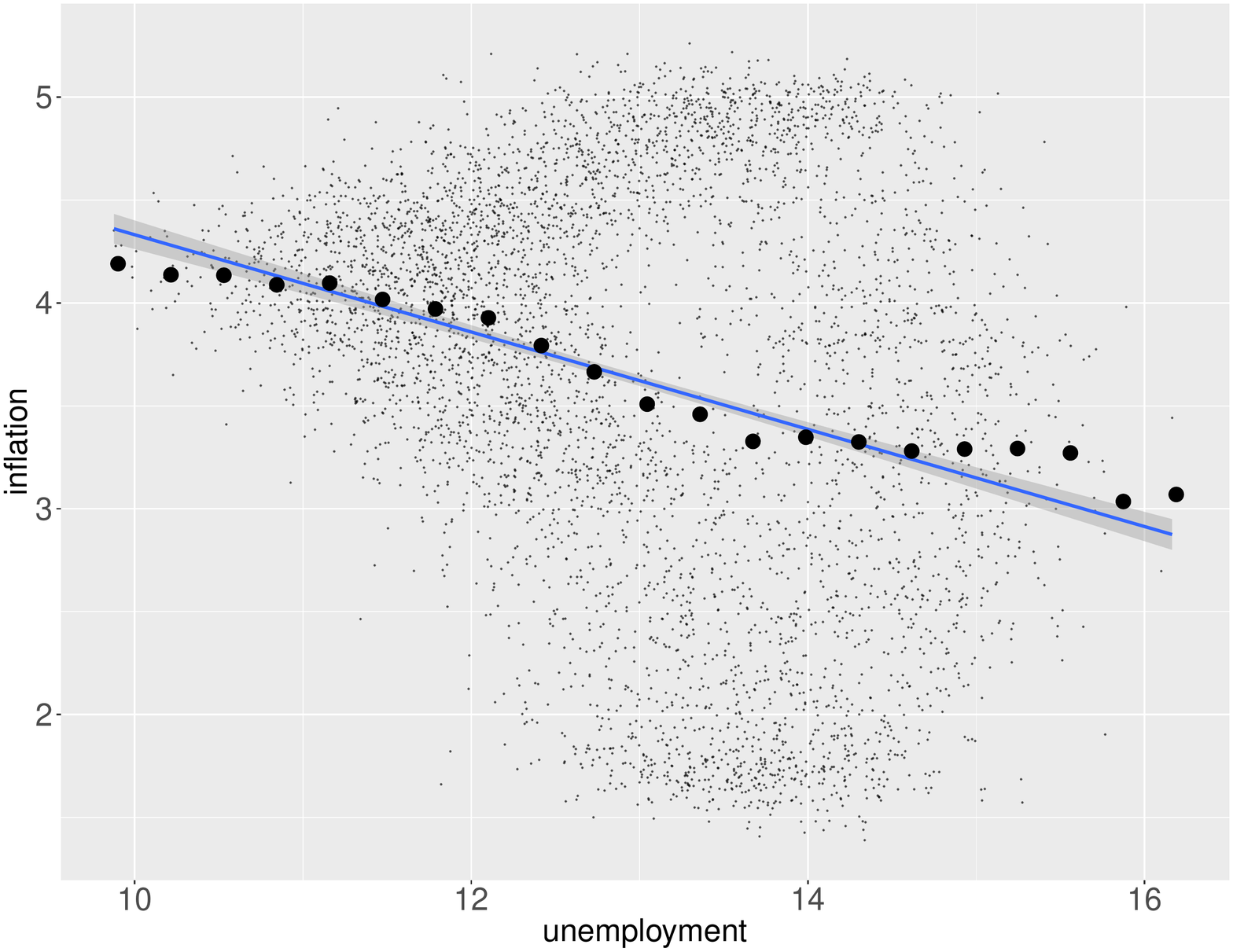}
\includegraphics[scale=0.3]{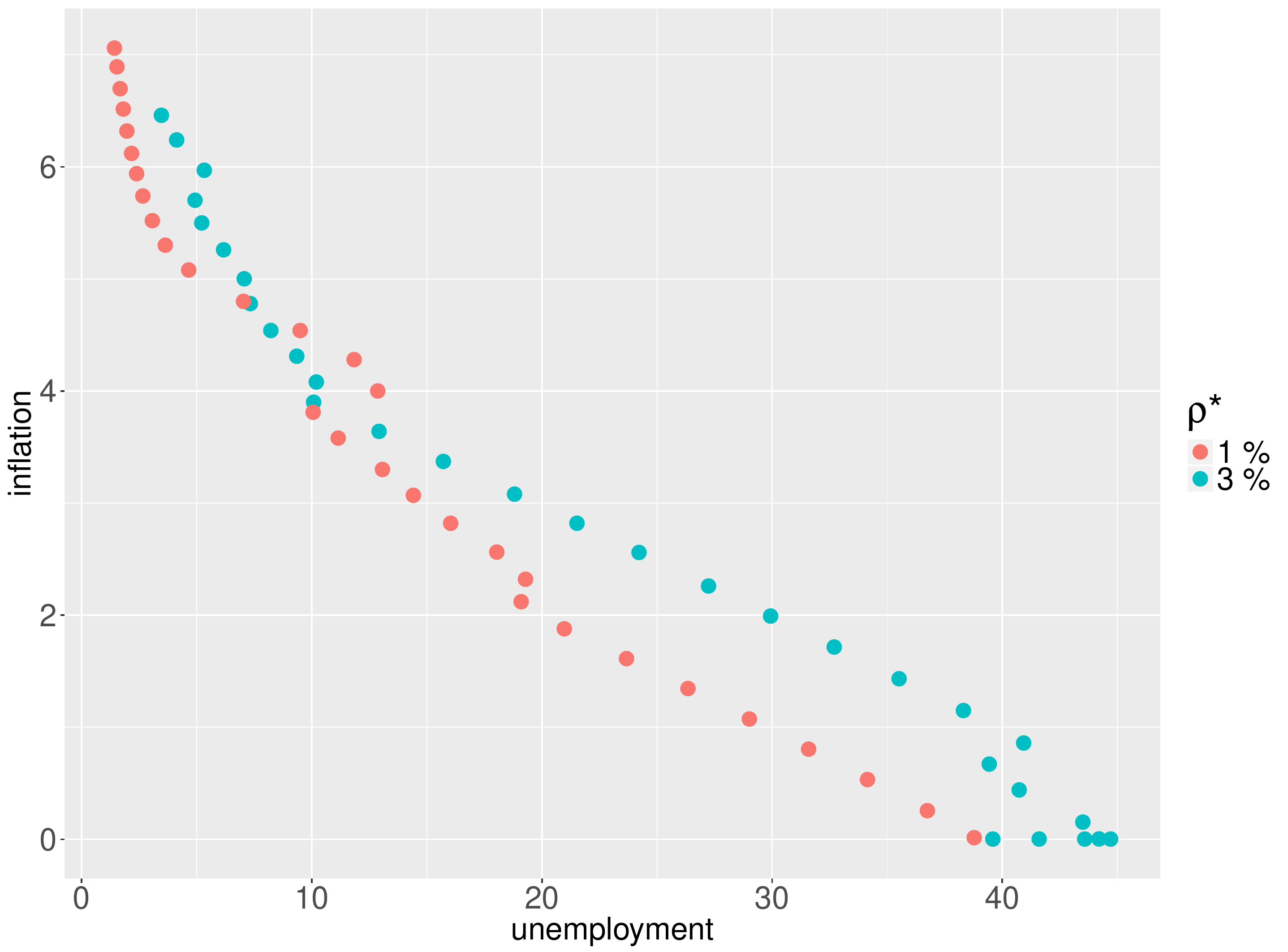}
\caption{Phillips curves, following two procedures: (Left) scatter plot of inflation vs. unemployement during the time evolution of the economy around a single equilibrium, obtained with 
$R=1.3$, $\rho^\star=3 \%$, $\phi_\pi=2.5$ and $\pi^\star=4 \%$, and $\tau^{\text{T}}=\tau^{\text{R}}=0.5$. 
(Right) scatter plot of the average inflation vs. average unemployment, extracted from panel b) and panel a) 
of Figs.~\ref{fig:targetHIHO} and~\ref{fig:targetLILO}, both
for $\phi_\pi=2.5$.}

\label{fig:PC}
\end{figure}

{
It is interesting to draw the Phillips curves predicted by our framework. There are two ways to think about these curves:
\begin{enumerate}
\item One is to draw realised inflation as a function of unemployment for
a fixed set of parameters, as the economy fluctuates over time around a unique equilibrium. 
\item The second is to represent the average realised inflation versus the average unemployement rate as the parameters of the economy, such as the
inflation target of the CB, are changed. This amounts to represent parametrically the results of panel b) versus those of panel a) in Figs.~\ref{fig:targetHIHO} and~\ref{fig:targetLILO}.
\end{enumerate}
In both cases we find a downward sloping relation (see Fig.~\ref{fig:PC}), compatible with the standard wisdom. Note however that the dynamical scatter plot obtained with procedure 1
leads to a noisy blob of points, with a downward sloping regression line. The second procedure leads to a nice looking graph, since each point is itself an average quantity. 
Reality should be in between, as one expects the underlying parameters of the economy to be themselves time dependent. But in any case, substantial deviations from a perfect downward sloping relation are expected.  
}

\subsection{Discussion}

The results of this section suggest that independently of the nature of its
native state, low inflation targets are detrimental to a CB-controlled
economy.\footnote{We should again insist that the numbers quoted should not
be taken at face value since no attempt has been made to calibrate Mark-0 on
real data. In particular, the chosen elementary time scale of 6-months is quite
arbitrary but directly scales all inflation and interest rates.} 
Interestingly, our results show that a situation where the realised inflation is
lower than the target inflation cannot be optimal; in fact, realised inflation
should rather {\it overshoot} target inflation on average (at least up to the point where  
savings are wiped out by inflation). This is the case for example in the LILO state discussed above, 
when unemployment reaches $8 \%$ for a target inflation of $4\%$, and a {realised} inflation of $5 \%$ 
(see Fig.~\ref{fig:targetLILO} for $\phi_\pi=2.5$).

Note that the coefficients $\tau^{\text{R}}$ and $\tau^{\text{T}}$, which here have been assumed to
be constant for simplicity, should in reality be time dependent and related to other quantities in the model.
For example, persistent inflation overshoots may result in a loss in the credibility of CB, which in the present
model means an increase of the value of $\tau^{\text{R}}$ and/or a decrease of $\tau^{\text{T}}$, 
as economic agents start looking for guidance in past realised inflation rather than in the CB target.

Such an increase can lead to a run-away inflation state (see Fig.~\ref{fig:native_ema}), which cannot be controlled anymore using Taylor-rule based policies. 
Within our model, such an hyper-inflation scenario can be tamed if firms do not fully index wages
on expected inflation (i.e. set the parameter $g$ 
to a value less than unity in Eq.~(\ref{eq:wages})). This has the effect of reducing 
realised inflation as households reduce consumption, pushing the hyper-inflation threshold 
$\tau^{\dagger}$ to higher values (for example, $\tau^{\dagger}
\approx 1.4$ for $g=0.5$). The smoothing parameter $\omega$ might itself depend on inflation,
as more volatility inflation could lead agents to become more short-sighted, such that $\omega \to 1$.
It would therefore be extremely interesting to extend the present model to include a dynamic coupling between $\tau^{\text{R}}$, 
$\tau^{\text{T}}$, and the target and realised inflation, as well as a dynamic dependence of $g$ and $\omega$ on inflation. We leave such a study for future work.

\section{Comparison with DSGE \& Conclusion} \label{sec:conclusions}

The issue of an optimal inflation target has only recently been considered
within the mainstream DSGE macroeconomic model~\cite{Ascari,Andrade}
{(see also~\cite{Grauwe} for a direct comparison of ABMs and DSGE models on a related topic).}
In this framework, the main cost
of inflation comes from 
price dispersion and is a consequence of the following string of
assumptions~\cite{Ascari}: a)~firms face friction costs and cannot update their
prices as often as they would like; 
b)~inflation leads to a stronger dispersion of (stale)
prices across different sectors of the economy; c)~stronger price deviations
from equilibrium lowers economic efficiency.  

\begin{figure}
\centering
\includegraphics[scale=0.3]{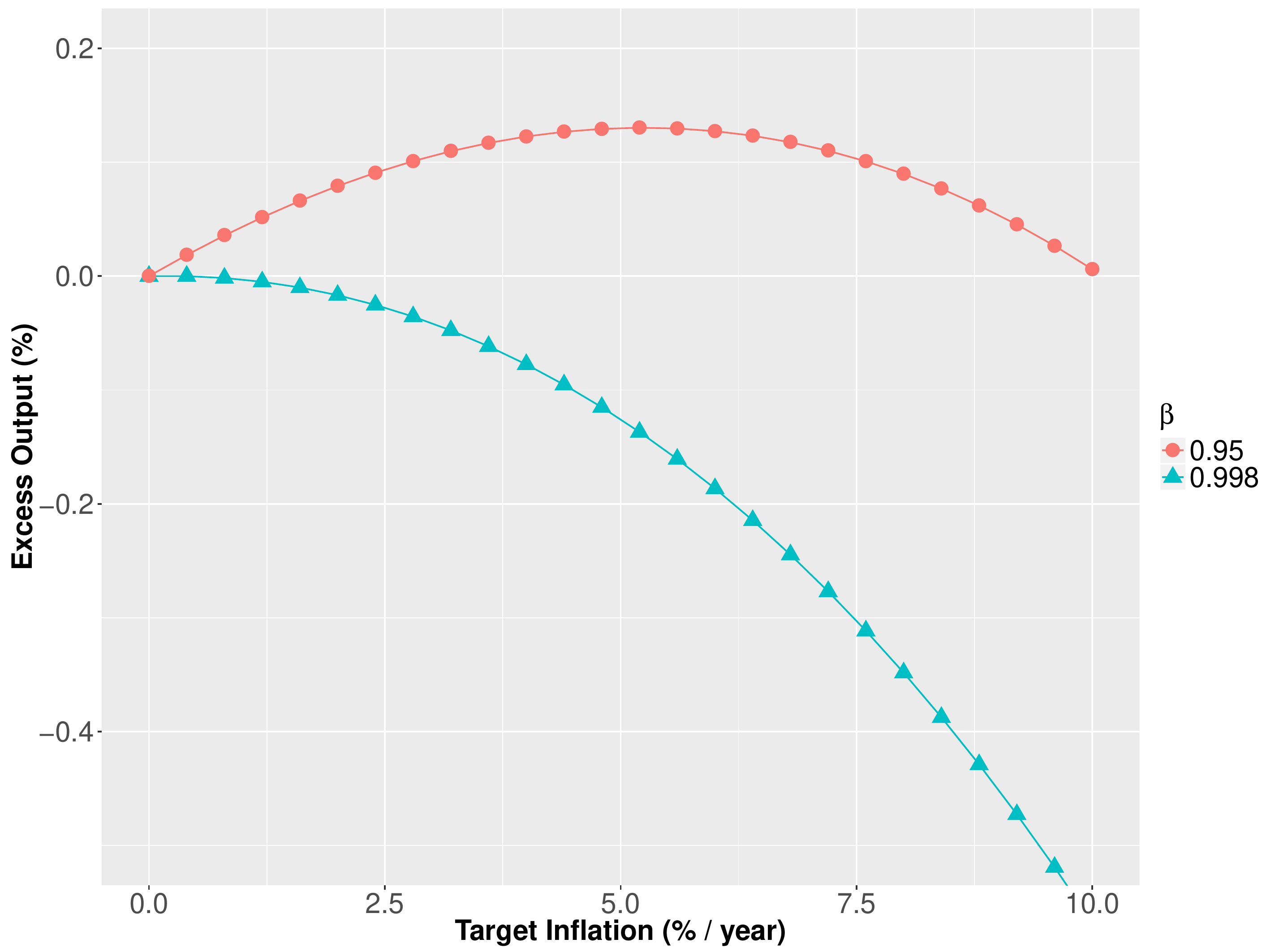}
\caption{Effect of inflation on output in the DSGE model,
following~\cite{Ascari2,Coibion}. All parameters are as in~\cite{Coibion}, but
the subjective discount factor $\beta$ is changed from 
$0.998$ (corresponding to a horizon of 125 years) to $0.95$ (corresponding to a horizon of 5 years). In the former case,
inflation causes output to decrease except for a very small window $\pi < 0.12
\%$/year, invisible in the graph. A shorter horizon leads to a positive marginal effect of the inflation on output as long as $\pi < 5.2 \%$.
}
\label{fig:dsge}
\end{figure}

However, while crucial in determining the optimal inflation rate within DSGE, 
such a dispersion induced cost has little empirical
support~\cite{Nakamura}. Embracing the choice of parameters and welfare function made by Coibion et
al.~\cite{Coibion}, 
the optimal inflation rate is found to be $1.5 \%$/year. This number is
highly dependent on the assumption made about the subjective discount factor
$\beta$ used by the representative 
household, i.e. how far in the future do economic agents assess the consequences
of their present decision. In many DSGE calibrations, the discounting horizon is
extremely long, for example 125 years (!) in Ref.~\cite{Coibion}. Although rooted in market efficiency arguments 
and based on the value of historical rates, such an enormous time scale is in our eyes
totally unreasonable. In line with the behavioral arguments used to construct
ABMs, where agents are assumed to be myopic, we believe that this time scale
should be rather on the scale of a few -- perhaps 5 -- years. This substantially
changes the conclusions of DSGE models, as the total output would then be an {\it
increasing} function of inflation up to $5.2 \%$, see Fig. \ref{fig:dsge}. 

The fact that a higher inflation tends to stabilize the Mark-0 economy is fully consistent with our previous results~\cite{Tipping,Policy}.
{Increasing the inflation rate discourages savings and motivates indebted firms to continue their business}, with the effect of lowering 
the unemployment rate and increasing global output. Hence, Mark-0 emphasizes the benefits of inflation while it completely neglects all inflation costs, 
including the price dispersion induced cost present (but probably overestimated) in DSGE models. 

More fundamentally, the most interesting difference between the DSGE and ABM modelling
strategies {is that the nature of native state} of the economy is {\it itself} an output of the ABM. This emergent 
state can change radically when the parameters characterizing the micro-behaviour of the different agents are only slightly modified. For example
the LILO, HIHO or hyper-inflation states considered in this paper are emergent properties of the
model, and not postulated a priori. Not surprisingly, a rough knowledge of where
the economy is ``naturally'' poised to go is needed to determine an adequate monetary policy. 
Trying to steer the economy too far from its native state is detrimental (as for the HIHO state) or even lead
to instabilities and crises (see~\cite{Policy}). {Quite interestingly, we have even found non-ergodic regions 
where the economy can be either in a LILO state or in a HIHO state depending on the initial conditions. This 
corresponds to potentially fragile situations where monetary policy could become extremely tricky.}

Mark-0 is a bare-bone ABM where many important effects are left out, that need to be considered in future studies. For example the network structure of
firms~\cite{Acemoglu,Bonart} and the feedback of investment on growth are clearly among the most urgent ingredients to be added in Mark-0. 
The difference with DSGE is that missing effects are straightforward to include in an ABM, 
while quite a bit of arm-twisting is usually necessary to include them in a DSGE
framework without ruining the mathematical tractability of the model. In this
sense, the much touted ``micro-founded'' nature of DSGE is quickly buried under a number of
ad-hoc assumptions (such as Calvo's sticky price mechanism~\cite{Calvo}), 
which are not much more convincing than the equally ad-hoc assumptions made in ABMs.  

In any case, the main result of our study is that the optimal inflation rate could be somewhat higher than the
currently accepted $2 \%$ target. One clear symptom of a too-low target is a persistent under-realisation of
inflation, perhaps similar to the current macroeconomic situation in the U.S. and in Europe. In our model, this 
predicament is alleviated by higher inflation targets that are found to improve 
both unemployment/output and negative interest rate episodes, up to the point where persistent over-realisation of inflation would lead to 
a loss of faith in the Central Bank and potential instabilities. 

Although our results are based on an arguably over-simplified model, it certainly militates for more work along these lines~\cite{Foley,Bookstaber}. After all,
DSGE models are themselves over-simplified and, as recently emphasized by O. Blanchard~\cite{Blanchard2,OREP}, {\it they have to
become less imperialistic and accept to share the scene with other approaches to
modelisation.}

\section*{Acknowledgments} 

The input and comments of O. Blanchard, R. Bookstaber, H. Dawid, D. Delli Gatti, D. Farmer, R. Farmer, X. Gabaix, C. Hommes and A. Kirman have been extremely useful. We also thank the anonymous referees and commentators of the first version of this paper, their comments have helped improving its quality.

\begin{appendix}
\section{Pseudo-code of Mark 0 with inflation expectations}
\label{app:Mark0}

We present here the pseudo-code for the Mark 0 code described in Sec.~\ref{sec:mark0recap}. 
The source code is available on demand.

\begin{algorithm}[h]
\caption{Mark 0}         
\label{alg:Mark0}                          
\begin{algorithmic} 
\Require$N_{\rm F} (10000)$; $c_0 (0.5)$; $\beta (2)$; $\gamma (0.1)$; $R$; $\eta^0_+ (R\eta^0_-)$; $\eta^0_- (0.2)$; $\delta (0.02)$;
$\Theta (3)$; 
$\varphi (0.1)$; $f (0.5)$; $\alpha_c(4)$; $\phi_\pi$; $\alpha_\Gamma(50)$; $\Gamma_0 (0)$;
$\pi^*$; $\rho^{\star}$; $\omega (0.2)$; $g (1)$; $\tau^{\text{R}}$; $\tau^{\text{T}}$; $T(10000)$.
Numbers between paretheses indicate the value used for the present work, the parameters with no default number have been varied in this work. 
We start computing averages
after $T_{eq} (5000)$ time steps.
\State
\Comment{ {\bf\color{red} Initialization}}
\State $Y_0 \leftarrow 0.1 + 0.9 \text{\tt random}$
\For {($i \leftarrow 0; i< N_{\rm F} ;i\leftarrow i+1$)}
\State $p[i] \leftarrow 1   + 0.1 (2\text{\tt random} - 1) $
\State $Y[i] \leftarrow Y_0 + 0.1 (2\text{\tt random} - 1)$
\State $D[i] \leftarrow Y_0$
\State $W[i] \leftarrow 1$
\Comment{Initial employment is random}
\State $\EE[i] \leftarrow 2 W[i] Y[i] \, \text{\tt random}$
\State $\PP[i] \leftarrow p[i] \min(D[i],Y[i]) - W[i] Y[i]$
\State $a[i] \leftarrow 1$
\Comment{binary variable: active ($1$) / inactive ($0$) firm}
\EndFor
\State $S\leftarrow N_{\rm F} - \sum_i \EE[i]$
\If{$\phi_\pi==0$}
\State $\pi^*  \leftarrow 0$
\State $\tau^T \leftarrow 0$
\EndIf
\Comment{ {\bf\color{red} Main loop}}
\For {($t \leftarrow 1; t \leq T;t\leftarrow t+1$)}  
\State $\varepsilon\leftarrow \frac1{ N_{\rm F}} \sum_i Y[i]$
\State $u \leftarrow 1-\varepsilon$
\State $\overline p \leftarrow \frac{\sum_i p[i] Y[i]}{\sum_i Y[i]}$ 
\State $\overline w \leftarrow \frac{\sum_i W[i] Y[i]}{\sum_i Y[i]}$ 
\State $u^*[i] \leftarrow \frac{\exp(\b W[i]/\overline w)}{\sum_i a[i]\exp(\b W[i]/\overline w)}N_{\rm F}u $
\State $x^{\text{ema}} \leftarrow \omega x +(1-\omega)x^{\text{ema}}$ where $x$ are $\pi,\rho^{\text{d}},\rho^{\ell},u$
\Comment{{\bf \color{blue} Central Bank policy}}
\State $\widehat{\pi} \leftarrow \tau^{\text{R}}\pi^{\text{ema}} + \tau^{\text{T}}\pi^*$
\State $\rho_0 \leftarrow \rho^{\star} + \phi_\pi(\pi^{\text{ema}}-\pi^*)$
\State $\G \leftarrow \max{\{\alpha_\Gamma(\rho^{\ell,\text{ema}} -
\widehat{\pi}),\Gamma_0 \}}$
\State $\mathcal{D} \leftarrow \EE^- \leftarrow \EE^+ \leftarrow 0$
\Comment{{\bf \color{blue} Firms update prices, productions and wages}}
\For {($i \leftarrow 0; i< N_{\rm F} ;i\leftarrow i+1$)}
\If { $a[i]==1$ }
\If {$\EE[i]>-\Theta W[i] Y[i]$}
\State $\EE^+\leftarrow \EE^++\max{\{\EE[i],0\}}$
\State $\EE^-\leftarrow \EE^- -\min{\{\EE[i],0\}}$
\State $\Phi[i] \leftarrow -\frac{\EE[i]}{W[i]Y[i]}$
\State $\h_+\leftarrow \qq{\h^0_+(1-\Gamma \Phi[i])}$
\State $\h_-\leftarrow \qq{\h^0_-(1+\Gamma \Phi[i])}$
\If { $Y[i] < D[i]$ }
\If {$\PP[i]>0$}
\State $W[i]\leftarrow W[i][1+\g(1-\Gamma \Phi[i])\varepsilon$ {\tt random}$]$
\State $W[i]\leftarrow \min{\{W[i],(P[i]\min{[D[i],Y[i]]} + \rho^{\text{d}}\max{\{\EE[i],0\}}+ \rho^{\ell}\min{\{\EE[i],0\}})/Y[i]\}}$
\EndIf
\State $Y[i] \leftarrow Y[i] + \min\{ \h_+  (D[i] - Y[i] ),  u^*[i] \}$
\If{ $p[i] < \overline p$} $p[i] \leftarrow p[i] ( 1 + \g \,${\tt random})
\EndIf
\ElsIf{ $Y[i] > D[i]$ }
\If {$\PP[i]<0$}
\State $W[i]\leftarrow W[i][1-\g(1+\Gamma \Phi[i])u$ {\tt random}$]$
\EndIf
\State $Y[i] \leftarrow \max\{ 0, Y[i] - \h_- ( D[i] - Y[i] ) \}$
\If{ $p[i] < \overline p$ } $p[i] \leftarrow p[i] ( 1 - \g \,${\tt random})
\EndIf
\EndIf
\State $p[i]\leftarrow p[i](1+\widehat{\pi})$
\State $W[i]\leftarrow W[i](1+g\widehat{\pi})$
\State $W[i]\leftarrow \max{(W[i],0)}$
\ElsIf {$\EE[i]\leq-\Theta W[i] Y[i]$}
\State $a[i]\leftarrow 0$
\State $\mathcal{D}\leftarrow  \mathcal{D}- \EE[i]$
\EndIf
\EndIf
\EndFor
      \algstore{Mark0}
  \end{algorithmic}
\end{algorithm}
\begin{algorithm}
  \caption{Mark0 (continued)}
  \begin{algorithmic}
      \algrestore{Mark0}
\State
\State $u \leftarrow 1-\frac1{ N_{\rm F}} \sum_i Y[i]$
\Comment{Update $u$ and $\overline p$}
\State $\overline p \leftarrow \frac{\sum_i p[i] Y[i]}{\sum_i Y[i]}$
\State
\Comment{{\bf \color{blue} Private bank sets interest rates}}
\State $\rho^{\ell} = \rho_0 + (1-f)\mathcal{D}/\EE^-$
\State $\rho^{\text{d}} = \frac{\rho^{\ell}\EE^--\mathcal{D}}{S+\EE^+}$
\State
\Comment{{\bf \color{blue} Households decide the demand}}
\State $S \leftarrow (1+\rho^{\text{d}})S + \sum_i W[i]Y[i]$
\State $c  \leftarrow c_0 [1+\alpha_c(\widehat{\pi}-\widetilde{\rho}^{\text{d,ema}})]$
\State $C_B \leftarrow c S$
\State
\For {($i \leftarrow 0; i< N_{\rm F} ;i\leftarrow i+1$)}
\State $D[i] \leftarrow \frac{C_B a[i] \exp(-\b p[i]/\overline p)}{p[i] \sum_i a[i]\exp(-\b p[i]/\overline p)} $
\Comment{Inactive firms have no demand}
\EndFor
\State
\Comment{{\bf \color{blue} Accounting}}
\State $\EE^+ \leftarrow 0$
\For {($i \leftarrow 0; i< N_{\rm F} ;i\leftarrow i+1$)}
\If { $a[i]==1$ }
\State $S \leftarrow S - p[i] \min\{ Y[i], D[i] \}$
\State $\PP[i] \leftarrow p[i] \min\{ Y[i], D[i] \} - W[i] Y[i]+ \rho^{\text{d}}\max{\{\EE[i],0\}}+ \rho^{\ell}\min{\{\EE[i],0\}}$
\State $\EE[i] \leftarrow \EE[i] + \PP[i]$
\If { $\PP[i] > 0$ \&\& $\EE[i]>0$ }
\Comment{Pay dividends}
\State $S \leftarrow S + \d \, \EE[i]$
\State $\EE[i] \leftarrow \EE[i] - \d \, \EE[i]$
\EndIf
\State $\EE^+\leftarrow \EE^++\max{\{\EE[i],0\}}$
\EndIf
\EndFor
\State
\Comment{{\bf \color{blue} Revivals}}
\State $\mathcal{R} \leftarrow 0$
\For {($i \leftarrow 0; i< N_{\rm F} ;i\leftarrow i+1$)}
\If { $a[i]==0$}
\If { {\tt random }$<\varphi$ }
\State $Y[i] \leftarrow u\mbox{ random}$
\State $a[i] \leftarrow 1$
\State $P[i] \leftarrow \overline p$
\State $W[i] \leftarrow \overline w$
\State $\EE[i] \leftarrow W[i]Y[i]$
\State $\mathcal{R}\leftarrow \mathcal{R}+\EE[i]$
\State $\EE^+\leftarrow \EE^++\max{\{\EE[i],0\}}$
\EndIf
\EndIf
\EndFor
\For {($i \leftarrow 0; i< N_{\rm F} ;i\leftarrow i+1$)}
\If{$a[i]==1$}
\If {$\EE[i]>0$}
\State $\EE[i] \leftarrow \EE[i] - \mathcal{R} \EE[i]/\EE^+$
\EndIf
\EndIf 
\EndFor
\EndFor  

\end{algorithmic}
\end{algorithm}

\end{appendix}

\end{document}